\documentclass[article]{IEEEtran}
\usepackage[intlimits]{amsmath}
\usepackage{amsfonts}
\usepackage{tikz}
\usepackage{cite}
 \usepackage[normalem]{ulem}
\usepackage{circuitikz}
 \usepackage{tkz-euclide}
 \usetikzlibrary{angles}
 \usetikzlibrary{positioning}
\usepackage{tikz-3dplot}
\usepackage[utf8]{inputenc}
\usepackage{color}
\usepackage{lipsum}
\usepackage{soul}
\usepackage{algorithm}
\usepackage{algpseudocode}
\usepackage{setspace}
\usepackage{float}
\usepackage{physics}
\newfloat{algorithm}{t}{lop}
\usepackage{physics}
\usepackage{multicol}
\usepackage{tabularx}
\usepackage{booktabs}

\usepackage[]{hyperref}
\hypersetup{
	colorlinks,
	linkcolor={red!50!black}, 
	citecolor={blue!50!black},
	urlcolor={blue!20!black}
}

\newcommand{\M}[1]{\mathbf{#1}}
\newcommand{\T}[1]{\mathrm{#1}}
\newcommand{\V}[1]{\boldsymbol{#1}}
\newcommand{\UV}[1]{\hat{\boldsymbol{#1}}}
\newcommand{\herm}{\T{H}}
\newcommand{\tran}{\T{T}}

\newcommand{\ie}{\textit{i}.\textit{e}.{}} 

\newcommand{\cf}{\textit{cf}.{}}

\newcommand{\fom}{\nu}

\title{Characteristic Modes of Nonreciprocal Structures}
\author{
Niklas Wingren, \IEEEmembership{Graduate Student Member, IEEE},
Daniel Sj\"oberg, \IEEEmembership{Senior Member, IEEE},\\ 
Mats Gustafsson, \IEEEmembership{Senior Member, IEEE},
Johan Lundgren, \IEEEmembership{Member, IEEE},\\
Miloslav Capek, \IEEEmembership{Senior Member, IEEE},
Lukas Jelinek, and
Kurt Schab, \IEEEmembership{Member, IEEE}

\thanks{Manuscript received \today; revised \today. This work was supported by the Czech Science Foundation under project~\mbox{No.~21-19025M}, and the Swedish Governmental Agency for Innovation Systems (Vinnova NFFP7-2023-01204).}%
\thanks{N. Wingren, D. Sj\"oberg, M. Gustafsson, and J. Lundgren are with Lund University, Lund, Sweden, (e-mails: \{niklas.wingren, daniel.sjoberg, mats.gustafsson, johan.lundgren\}@eit.lth.se).}%
\thanks{L. Jelinek and M. Capek are with Czech Technical University in Prague, Czech Republic (e-mails: \{lukas.jelinek,miloslav.capek,\}@fel.cvut.cz).}
\thanks{K. Schab is with Santa Clara University, Santa Clara, USA (e-mail: kschab@scu.edu).}%
}
\begin{document}
\let\Algorithm\algorithm
\renewcommand\algorithm[1][]{\Algorithm[#1]\setstretch{1.4}}

\maketitle

\begin{abstract}
The scattering formulation of characteristic mode decomposition is utilized to extend modal analysis to lossless scatterers breaking time-reversal symmetry. This enables characteristic modes analysis on devices containing gyrotropic or moving media.  The resulting nonreciprocity introduces features not observed in reciprocal scenarios, such as asymmetric phase progression in characteristic far fields. These new phenomena are carefully discussed using examples of varying complexity.  Indicators of nonreciprocity based on modal data are also introduced.
\end{abstract}


\section{Introduction}

\IEEEPARstart{C}{haracteristic} modes are an intensively studied branch of classical electrodynamics with applications in antenna and scattering theory~\cite{2022_Lau_APM}. Commonly, characteristic mode analysis is associated with the method of moments and the study of perfectly conducting bodies~\cite{HarringtonMautz_TheoryOfCharacteristicModesForConductingBodies,capek2022computational}, but its original scattering formulation~\cite{1948_Montgomery_Principles_of_Microwave_Circuits,Garbacz_TCMdissertation}, recently expanded and formalized in~\cite{Gustafsson_etal2021_CMAT_Part1,Gustafsson_etal2021_CMAT_Part2,Capek+etal2022}, significantly broadens the application of characteristic modes to scatterers comprised of arbitrary materials. This immediately raises questions about the properties of characteristic modes in the presence of exotic materials, such as those with bianisotropic response and/or those in which nonreciprocity stems from breaking time-reversal symmetry~\cite{2014_Fleury_Science}. These scenarios emerge when studying antennas on gyrotropic substrates~\cite{1992_Pozar_TAP,2009_Ozgur_JMSME1,2009_Ozgur_JMSME2} or material bodies in motion~\cite{van2012relativity}, and give characteristic modes additional parameterization in quantities such as the intensity of a biasing field or velocity.  This letter initiates a discussion on the characteristic modes of lossless nonreciprocal scatterers by showing their fundamental properties and examining indicators of modal nonreciprocity.

\section{Scattering-based characteristic modes}

This paper uses a formulation of characteristic modes based on an object's scattering matrix~\cite{1948_Montgomery_Principles_of_Microwave_Circuits} recently formalized in \cite{Gustafsson_etal2021_CMAT_Part1,Gustafsson_etal2021_CMAT_Part2,Capek+etal2022}. In contrast to impedance formulations~\cite{HarringtonMautz_TheoryOfCharacteristicModesForConductingBodies,HarringtonMautz_ComputationOfCharacteristicModesForConductingBodies}, the scattering approach avoids problems reported in~\cite{Gustafsson_etal2021_CMAT_Part1} in cases where the radiation matrix is not equal to the real (Hermitian) part of the impedance matrix, see \cite[Eqs. (10)--(11)]{HarringtonMautz_TheoryOfCharacteristicModesForConductingBodies}.

Using the scattering matrix~$\M{S}$ of a linear material object~\cite{1948_Montgomery_Principles_of_Microwave_Circuits}, characteristic modes are defined by the eigenvalue problem~\cite{Garbacz_1965_TCM,Gustafsson_etal2021_CMAT_Part1}
\begin{equation}
    \M{S}\M{a}_n = s_n\M{a}_n,
    \label{eq:CMdef}
\end{equation}
where the complex eigenvalue~$s_n = 1 + 2 t_n$ is commonly parametrized by a real characteristic number~$\lambda_n$ via relation \mbox{$s_n = -(1 - \T{j} \lambda_n)/(1 + \T{j} \lambda_n)$} and where eigenvectors~$\M{a}_n$ collect the expansion coefficients of the fields in the selected basis, such as spherical vector waves commonly used in the case of finite scatterers in free space~\cite{Kristensson_ScatteringBook}. The modal significance of an eigenvector is denoted as \mbox{$\left|t_n \right| = \left| 1 + \T{j} \lambda_n \right|^{-1} = \left|s_n - 1\right|/2$}. Other characteristic quantities, such as the characteristic currents, can be computed by exciting the system with the vector~$\M{a}_n$ and evaluating the desired quantity~\cite{Gustafsson_etal2021_CMAT_Part1}.

For lossless scatterers, the scattering matrix is unitary,~$\M{S}^\herm \M{S} = \M{1}$, and pairs of characteristic excitations can be made orthonormal, \ie{},
\begin{equation}
    \M{a}_m^\herm \M{a}_n = \delta_{mn},
    \label{eq:otho}
\end{equation}
where~$\delta_{mn}$ is the Kronecker delta and~$^\herm$ denotes Hermitian transpose.

\section{Characterizing Reciprocity and Nonreciprocity}
Characteristic modes of a lossless scatterer satisfy~\cite[App.~ B]{Gustafsson_etal2021_CMAT_Part1}
\begin{equation}
    \M{S}^\T{T}\M{a}_n^* = s_n\M{a}_n^*
    \label{eq:CMdef1}
\end{equation}
due to the unitarity of~$\M{S}$. The scattering matrix is symmetric $\M{S}^\T{T}=\M{S}$ for reciprocal scatterers, in which case both $\M{a}_n$ and $\M{a}_n^*$ are eigenvectors of $\M{S}$ with the same eigenvalue. Hence, the eigenvectors can always be made real-valued\footnote{Eigenvectors for non-degenerate cases are generally equiphase (with orthonormal $\M{a}_n$ satisfying $|\M{a}_n^{\T{T}}\M{a}_n|=1$). For a degenerate case, the modes can exhibit phase progression, but linear combinations of vectors~$\M{a}_n$ and~$\M{a}_n^*$ can be used to produce real-valued eigenvectors.}, akin to the treatment of characteristic currents~\cite{HarringtonMautz_TheoryOfCharacteristicModesForConductingBodies}.  Due to symmetries of vector spherical waves~\cite[Sec.~7.8]{Kristensson_ScatteringBook}, the real-valued characteristic vectors~$\M{a}_n$ imply conjugate inversion symmetry~\cite[Eqs.~(29)--(30)]{Kahn_MSA_1965} of the characteristic far fields
\begin{equation}
    \V{F}_n(-\hat{\V{r}})=-\V{F}_n^{\ast}(\hat{\V{r}}),
    \label{eq:Fsym}
\end{equation}
where $\hat{\V{r}}$ is an observation direction $\hat{\V{r}}=\V{r}/r$ and the far field $\V{F}$ is defined by the scattered field $\V{E}_\T{s}$ as \mbox{$\V{F}(\hat{\V{r}}) = r\,\T{exp}\left\{\T{j}kr\right\}\V{E}_\T{s} (\V{r})$} for $r\to\infty$.
This property also follows from selecting real-valued currents in reciprocal EFIE formulations of characteristic modes\footnote{The integral relationship between a real-valued current distribution and its radiated far-field~\cite[\S 3.5-3.6]{Balanis_Wiley_2005} illustrates this property.}. 
 
The above-mentioned properties of reciprocal systems (real-valued eigenvectors, far-field inversion symmetry) are lost in nonreciprocal systems where $\M{S}\neq\M{S}^\T{T}$, and their absence can be utilized as distinguishing features to characterize nonreciprocity. To detect nonreciprocity of characteristic modes, the scattering matrix~$\M{S}$ is decomposed into its symmetric and antisymmetric parts as $\M{S} = (\M{S} + \M{S}^\tran)/2 + (\M{S} - \M{S}^\tran)/2$.  This implies that the cycle-mean outgoing power~$P_\T{o}$ can be written as
\begin{multline}
    P_{\T{o}} = \dfrac{1}{2} |\M{S}\M{a}|^{2} = \dfrac{1}{8} \left|(\M{S} + \M{S}^\tran)\M{a}\right|^{2} \\
    + \dfrac{1}{8} \left|(\M{S} - \M{S}^\tran)\M{a}\right|^{2} + \dfrac{1}{4}\left( \left|\M{S}\M{a}\right|^{2} - \left|\M{S}^\tran\M{a}\right|^{2} \right).
\end{multline}
We denote the first term on the right-hand side as the reciprocal outgoing power $P_{\T{o}}^\T{r}$, while the second row corresponds to the nonreciprocal outgoing power $P_{\T{o}}^{\T{nr}}$. We then propose the ratio 
\begin{equation}
    \fom = \dfrac{P_{\T{o}}^{\T{nr}}}{P_{\T{o}}} \in \left[0 , 1 \right]
    \label{eq:fom}
\end{equation} 
to characterize the presence of nonreciprocity for a given excitation at a particular frequency. Furthermore, modal far fields~$\V{F}_n$ violating~\eqref{eq:Fsym}, such as those with asymmetric phase progression, are an indicator of nonreciprocal behavior. Together, the study of the quantity~$\fom$ and modal phase progression serve to classify individual characteristic modes as either reciprocal or nonreciprocal. Another classification of nonreciprocity based on the extinction cross section has recently been proposed in~\cite{nefedkin2023nonreciprocal}, but is not readily applicable to characteristic modes.

\section{Examples}

\subsection{Ideal circulator}
\label{sec:circ}

As an initial example demonstrating characteristic modes for nonreciprocal systems, we consider the scattering matrix of an ideal $N$-port circulator~\cite[\S 7.1]{Pozar_MicrowaveEngineering}
\begin{equation}
	\M{S} = \begin{bmatrix}
		0 & 1 & 0 & \cdots & 0\\
		0 & 0 & 1 & \cdots & 0\\
		\vdots & \vdots & \vdots & \ddots & \vdots \\
		0 & 0 & 0 & \cdots & 1\\
		1 & 0 & 0 & \cdots & 0
	\end{bmatrix}.
 \label{eq:s-circ}
\end{equation}
The characteristic modes (in this case, eigenmodes of the matrix $\M{S}$~\cite{Gustafsson_etal2021_CMAT_Part1}) associated with this system~\cite{auld1959synthesis} can be classified as progressive phase modes
\begin{equation}
    \begin{aligned}
        \M{a}_{\pm p} &= N^{-1/2} \left[1,\T{e}^{\pm\T{j}2p\pi/N},\cdots, \T{e}^{\pm\T{j}2p\pi(N-1)/N}\right]^\T{T}, \\
        s_{\pm p} &= \T{e}^{\pm\T{j}2p\pi/N}, \quad 1\leq p \leq (N-1)/2,
    \end{aligned}
\end{equation}
with $|\M{a}_{\pm p}^{\T{T}}\M{a}_{\pm p}|=0$, an equiphase even mode
\begin{equation}
	\M{a}_\T{e} = N^{-1/2}\left[1,1,\cdots,1\right]^\T{T},\quad s_\T{e} = 1,
\end{equation}
with $|\M{a}_\T{e}^{\T{T}}\M{a}_\T{e}|=1$, and an equiphase odd mode (even $N$ only)
\begin{equation}
	\M{a}_\T{o} = N^{-1/2}\left[1,-1,\cdots,1,-1\right]^\T{T},\quad s_\T{o} = -1
\end{equation}
with $|\M{a}_\T{o}^{\T{T}}\M{a}_\T{o}|=1$. 
 Application of the definition in \eqref{eq:fom} to each class of characteristic modes gives
\begin{equation}
	\fom_{\pm p} = \sin^2(2p\pi/N),\quad \fom_\T{e} = 0,\quad \fom_\T{o} = 0.
 \label{eq:circulator-fom}
\end{equation}
Similarly, the modal significance of the modes reads
\begin{equation}
	\left| t_{\pm p} \right| = \left| \sin(p\pi/N) \right|,\quad \left| t_\T{e} \right| = 0,\quad \left| t_\T{o} \right| = 1.
 \label{eq:circulator-MS}
\end{equation}
These results show that characteristic modes of nonreciprocal devices (such as this ideal circulator) can exhibit a range of values of the parameter~$\fom$, including particular excitations which, to the outside observer, cause the system to behave in a completely reciprocal manner with $\fom = 0$, \cf{}, the equiphase even / odd modes in \eqref{eq:circulator-fom}. In contrast, the progressive phase modes $\M{a}_{\pm p}$ cannot be normalized to be real-valued and exhibit non-zero values~$\nu$, both indicators of non-reciprocal behavior.

Scattering or transition matrices analogous to~\eqref{eq:s-circ} can be used to describe hypothetical objects exhibiting nonreciprocal coupling between the incident and scattered spherical waves, and a scattering dyadic can be constructed to describe non-reciprocal scattering on a direction-by-direction basis~\cite{Capek+etal2022}.

\subsection{Nonreciprocal subcircuit}

\begin{figure}
    \centering
  \includegraphics[clip,width=3.25in]{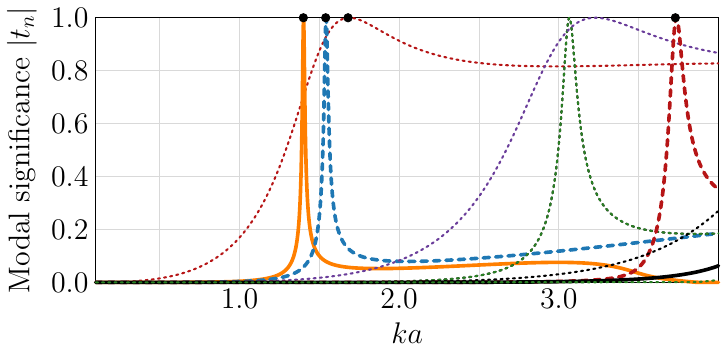}
   \begin{tikzpicture}\node at (0,0) {\includegraphics[width=3.25in]{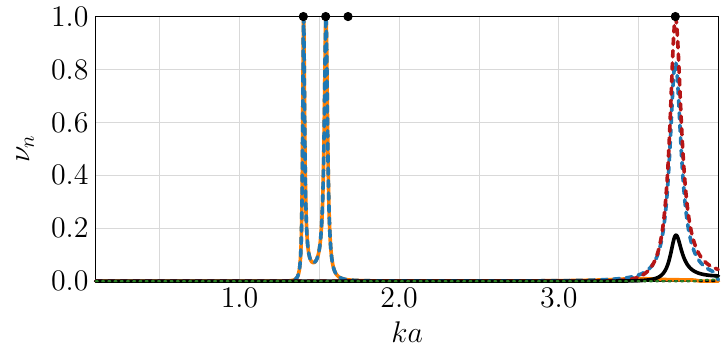}};
    \node[scale=0.23] at (-2.2,0.30) {\includegraphics{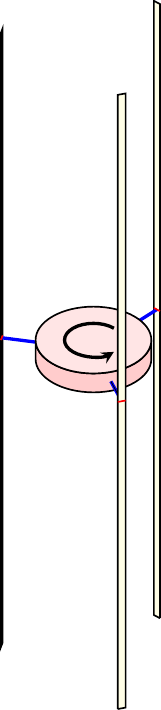}};
    \end{tikzpicture}
    \caption{(Top) Modal significances of three parallel dipoles connected to an ideal circulator. (Bottom) Nonreciprocity ratio~$\fom$ corresponding to modes depicted in the top panel.}
    \label{fig:dipolePlusCirculator}
\end{figure}

The second example deals with a free-space scatterer in which nonreciprocity is induced by the presence of an ideal lumped circulator. Specifically, the circulator is connected to delta-gap ports of three thin-strip dipoles of length~$L$ and width~$L/50$, see the inset in~Fig.~\ref{fig:dipolePlusCirculator}. The dipoles are made of a perfect electric conductor, are parallel to the~$z$-axis, and are positioned so that their centers lie on the circle with radius~$L/8$. The dipoles are distributed uniformly along this circle. The circulator is placed at the origin and is connected with the dipole ports via radial transmission lines of length~$L/8$ and characteristic impedance $Z_0 = 50 \, \T{\Omega}$ matching the system impedance of the ideal circulator. The scattering scenario is modeled using a combination of the method-of-moments~\cite{atom} and the scattering matrix description of a three-port circuit~\cite{Pozar_MicrowaveEngineering}.

Modal significances of this system are depicted in the top panel of Fig.~\ref{fig:dipolePlusCirculator}, while the nonreciprocity ratio $\fom$ is shown in the bottom panel of Fig.~\ref{fig:dipolePlusCirculator}. The color coding of both figures is identical. Most of the modal spectrum shows negligible values of the ratio~$\fom$. Exceptions are peaks at $ka \approx \left\{1.40, 1.54, 3.73 \right\}$, see black circle markers at the top of the plots, also exhibiting high modal significance in the same modes. The other modes with high modal significance, such as those belonging to the red dotted line in Fig.~\ref{fig:dipolePlusCirculator}, show reciprocal behavior with $\fom = 0$ and symmetric equiphase radiation patterns satisfying~\eqref{eq:Fsym}.

\begin{figure}
    \centering
    \includegraphics[width=3.25in]{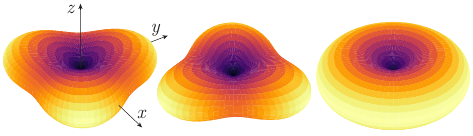}
    \caption{Directivity patterns corresponding to solid orange peak at $ka \approx 1.40$, dashed blue peak at $ka \approx 1.54$, and dotted red peak at $ka \approx 1.68$ in Fig.~\ref{fig:dipolePlusCirculator},  see black circle markers at the top of the plot.}
    \label{fig:DpattdipolePlusCirculator}
\end{figure}

\begin{figure}
    \centering
    \includegraphics[width=3.25in]{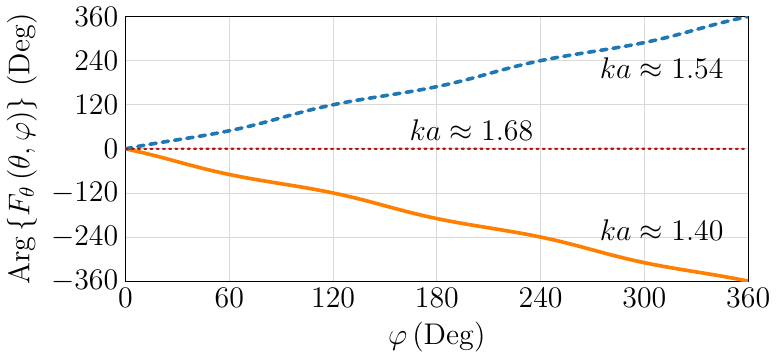}
    \caption{Phase of electric far-field corresponding to Fig.~\ref{fig:DpattdipolePlusCirculator}. Only $\theta$-component in the $xy$ plane is shown.}
    \label{fig:PhasePattdipolePlusCirculator}
\end{figure}

The nonreciprocal effects associated with the two peaks at $ka \approx 1.40$ and $ka \approx 1.54$ are best observed in the modal far fields, which are depicted in Fig.~\ref{fig:DpattdipolePlusCirculator} as a directivity pattern and in Fig.~\ref{fig:PhasePattdipolePlusCirculator} as a phase of the $\theta$ component of the electric far field in the $xy$-plane ($\theta = \pi/2$). As a reference, the right panel in Fig.~\ref{fig:DpattdipolePlusCirculator} shows the radiation pattern of an omnidirectional (inversely symmetric) significant reciprocal mode at $ka \approx 1.68$. The dashed red trace in Fig.~\ref{fig:PhasePattdipolePlusCirculator} also shows that this radiation pattern is equiphase. It can be seen that both nonreciprocal examples share features which contrast with the reciprocal mode.  Notably, their radiation diagrams lack inversion symmetry~\eqref{eq:Fsym} and exhibit $120^\circ$ rotational $\T{C}_{3z}$ symmetry and phase progression along angular direction $\varphi$. These modal radiation patterns represent field rotation around the $z$-axis, a property that is not present around reciprocal scatterers.  Note that because of the $\T{C}_{3z}$ symmetry of this particular system, its characteristic modes bear a strong resemblance to the $p=\pm 1$ modes of the ideal circulator discussed in Sec.~\ref{sec:circ}, with $120^\circ$ phase offsets between radiated fields at locations separated by that same angular separation in the $xy$ plane.

\subsection{General nonreciprocal media}
The last example presents characteristic modes of a penetrable body made of a nonreciprocal medium, showing the generality of the scattering-based formulation of characteristic modes for systems not readily analyzed using impedance-based methods. Consider a cylinder of radius~$\rho$ and height~$h$, centered at the origin with its axis lying along axis~$z$, filled with a homogeneous, real-valued permittivity~$\varepsilon_\T{r}\varepsilon_0$ and permeability~$\mu_\T{r}\mu_0$. Let the cylinder rotate around its axis with angular frequency~$\omega$ such that the velocity of a point of the cylinder reads
$\V{v} = \omega \left( \UV{y} x - \UV{x} y \right)$. The maximum speed is assumed to be much smaller than the speed of light~$c_0 = 1/\sqrt{\mu_0\varepsilon_0}$, \ie~$\beta_\rho = \omega \rho / c_0 \ll 1$.

The unidirectional movement of the cylinder breaks time-reversal symmetry and introduces nonreciprocity analogous to the motion of fluid studied in~\cite{2014_Fleury_Science, 2023_Wang_PRB}. To study the electromagnetic scattering from such an object and to evaluate its characteristic modes, the rotating cylinder can be replaced by a stationary cylinder with 
constitutive relations~\cite{van2012relativity,van1976electromagnetic}
\begin{equation}
\begin{aligned}
\V{D} &= \V{\varepsilon} \cdot \V{E} + c_0^{-1}\V{\xi } \cdot \V{H}\\
\V{B} &= c_0^{-1}\V{\varsigma} \cdot \V{E} + \V{ \mu } \cdot \V{H}
\end{aligned}
\label{eq:const1}
\end{equation}
where $\V{\varepsilon } = \varepsilon_\T{r}\varepsilon_0 \M{1}$ and $\V{\mu } = \mu_\T{r}\mu_0 \M{1}$
% \begin{equation}
% \V{\varepsilon } = \varepsilon \M{I}, \, \V{\mu } = \mu \M{I}
% \end{equation}
with $\M{1}$ denoting the identity tensor and (using the Cartesian coordinate system)
\begin{equation}
\V{\xi } = \omega \dfrac{\varepsilon_\T{r}\mu_\T{r} - 1}{c_0} 
\mqty(
0 & 0 & x \\
0 & 0 & y \\
-x & - y & 0
), \quad \V{\varsigma } =  - \V{\xi }.
\label{eq:const2}
\end{equation}
The constitutive relations above describe a lossless ($\V{ \varepsilon } = \V{ \varepsilon }^\herm$, $\V{ \mu } = \V{ \mu }^\herm$, $\V{ \varsigma } =  \V{ \xi }^\herm$), but inhomogeneous and nonreciprocal ($\V{ \varsigma } \ne  - \V{ \xi }^\T{T}$) medium~\cite{Kristensson_ScatteringBook}.

A common benchmarking example for characteristic modes is a dielectric cylinder of radius~\mbox{$\rho = 5.25\,\T{mm}$} and height~\mbox{$h = 4.6\,\T{mm}$} with \mbox{$\varepsilon_\T{r} = 38$}, and \mbox{$\mu_\T{r} = 1$} \cite{Gustafsson_etal2021_CMAT_Part2}. Applying the transformation in \eqref{eq:const1}--\eqref{eq:const2}, the characteristic modes of this system are computed for various relative speeds \mbox{$\beta_\rho \leq 0.01$} using the iterative method~\cite{Lundgren+etal2022}, which employs a finite element-boundary integral hybrid solver capable of computations involving general bianisotropic media~\cite{Wingren2023github}. Speeds of this order have been shown to significantly impact the scattering characteristics compared to the stationary case~\cite{van1976electromagnetic}.

\begin{figure}
    \centering
    \includegraphics[width=3.25in]{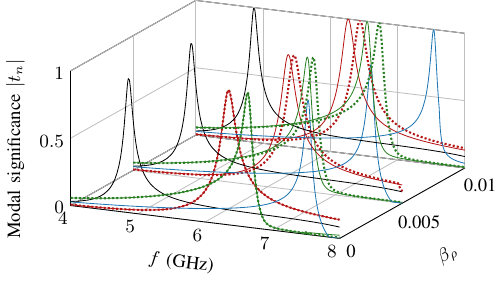}
    \caption{Modal significances of a rotating cylinder for different $\beta_\rho$. The first 6 modes are displayed, corresponding to TE$_{01}$ (black), HEM$_{11}$ (red), HEM$_{12}$ (green) and TM$_{01}$ (blue). Both HEM modes are degenerated at $\beta_\rho = 0$.}
    \label{fig:RotatingCylinderVariation}
\end{figure}

Modal significances of the cylinder at three different~$\beta_\rho$ are shown in Fig.~\ref{fig:RotatingCylinderVariation}. The resonant frequencies of the modes at different~\mbox{$\beta_\rho$} are shown in Table~\ref{tab:RotatingCylinderPeaks} for clarity. For the stationary case ($\beta_\rho=0$), the results match those in~\cite{Gustafsson_etal2021_CMAT_Part2}. There are six modes displayed, two of which are rotationally symmetric (TE$_{01}$ and TM$_{01}$) and four of which are degenerate pairs (HEM$_{11}$ and HEM$_{12}$). All modes have $\fom = 0$ and exhibit symmetric far fields satisfying~\eqref{eq:Fsym}. As rotation increases ($\beta_\rho = 0.005$ and $\beta_\rho = 0.01$), the symmetric TE$_{01}$ and TM$_{01}$ modes are relatively unaffected and preserve their symmetry properties, as their symmetry axis matches the axis of rotation. In contrast, the mode pairs which are degenerate at $\beta_\rho=0$ show level splitting with their frequency separation increasing for higher~$\beta_\rho$, similarly to the case studied in~\cite{2014_Fleury_Science}.

\begin{figure}
    \centering
    \includegraphics[width=3.25in]{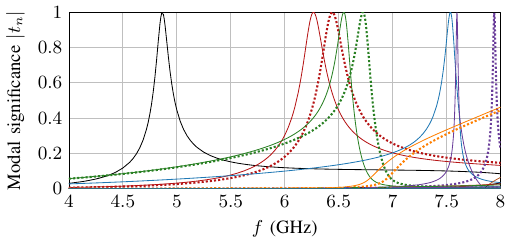}
    \includegraphics[width=3.25in]{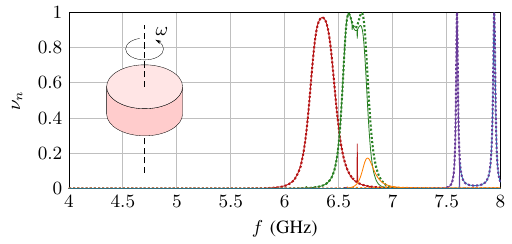}
    \caption{(Top) Modal significance of a rotating cylinder with $\beta_\rho=0.01$. (Bottom) Nonreciprocity ratio~$\fom$ corresponding to modes depicted in the top panel. Solid/dashed lines of the same color are degenerate modes at $\beta_\rho=0$.}
    \label{fig:RotatingCylinder}
\end{figure}

For $\beta_\rho=0.01$, modal significances and $\fom$ are shown in Fig.~\ref{fig:RotatingCylinder} for the 10 first modes. It is seen that non-zero $\fom$ is present in all modes which exhibit level splitting. It is, however, also seen that the mode drawn as a blue line has significant $\fom$ at~\mbox{$f > 7.62\,\T{GHz}$}. The modal significance of this mode matches the TM$_{01}$ at~\mbox{$f < 7.62\,\T{GHz}$}, but is swapped with the solid purple line at~\mbox{$f > 7.62\,\T{GHz}$} due to crossing avoidance, which occurs in nonsymmetric meshes~\cite{SchabBernhard_GroupTheoryForCMA}. This forces the far field of the blue line to abruptly change from rotationally symmetric to non-symmetric, which is accompanied by a similarly abrupt change in $\fom$. The spikes in~$\fom$ for the solid green and red lines at~\mbox{$f = 6.67\,\T{GHz}$} are also related to the tracking as they correspond to a point where crossing avoidance does not happen due to coarse frequency resolution. Together, these observations show how the characterization of nonreciprocity in characteristic modes is sensitive to tracking.

\begin{table}[]
    \centering
    \caption{Resonant frequencies in GHz of modes in Fig.~\ref{fig:RotatingCylinderVariation} at different $\beta_\rho$.}
    \begin{tabular}{ccccccc} \toprule
       $\beta_\rho$ & TE$_{01}$ & \multicolumn{2}{c}{HEM$_{11}$} & \multicolumn{2}{c}{HEM$_{12}$} & TM$_{10}$\\\midrule
       0  & 4.86 & 6.35 & 6.35 & 6.63 & 6.63 & 7.53 \\
       0.005  & 4.86 & 6.30 & 6.39 & 6.59 & 6.68 & 7.53 \\
       0.01 & 4.86 & 6.27 & 6.44 & 6.55 & 6.73 & 7.54 \\ \bottomrule
    \end{tabular}
    \label{tab:RotatingCylinderPeaks}
\end{table}

\begin{figure}
    \centering
    \includegraphics{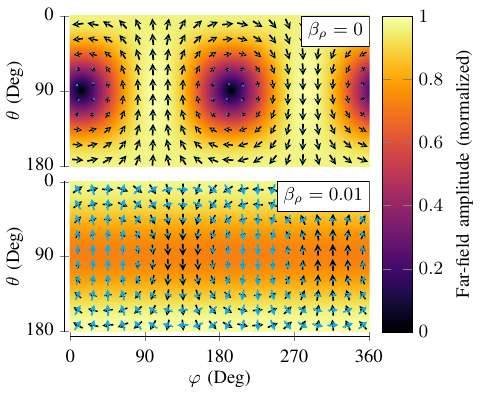}
    \caption{Far field for the HEM$_{11}$ mode corresponding to the solid red line in Fig.~\ref{fig:RotatingCylinderVariation} at two different $\beta_\rho$. Color images show the amplitude of the far field normalized to its peak value, while arrows indicate the directions of real (black) and imaginary (blue) parts.}
    \label{fig:RotatingCylinderFarfield}
\end{figure}

The far field corresponding to one of the nonreciprocal modes is illustrated in Fig.~\ref{fig:RotatingCylinderFarfield} for the cylinder stationary and rotating with $\beta_\rho=0.01$. It is seen that the far field can be selected to be purely real with inversion symmetry~\eqref{eq:Fsym} in the stationary case, but not in the rotating case. In the stationary case, the HEM mode is degenerated with the far field of the other degenerate mode being shifted $90^\circ$ in the $\varphi$ coordinate. While the two far fields in Fig.~\ref{fig:RotatingCylinderFarfield} look very different, it is noted that a far field very similar to that at $\beta_\rho=0.01$ can be obtained by linearly combining the two degenerate modes at $\beta_\rho=0$ with one mode phase shifted $\pi/2$.

\section{Discussion and Conclusions}

This letter describes how characteristic mode analysis can be applied to nonreciprocal devices, thereby extending the established theoretical and computational state-of-the-art. This advance enables the analysis of problems involving gyrotropic materials, moving media, and lumped non-reciprocal devices. The decomposition employs scattering matrices, which can be computed numerically in arbitrary full-wave solvers. Certain cases, such as a spherical shell filled with bianisotropic material, even retain the potential for analytical solutions~\cite{Kristensson_ScatteringBook}.

Examples discussed throughout the letter illustrate how specific modal properties are lost when a system exhibits nonreciprocal behavior. Modal nonreciprocity varies with frequency, hindering straightforward single-frequency classification of a mode as being (non)reciprocal. One remaining challenge lies in defining a measure that effectively quantifies modal nonreciprocity.

\end{document}